# Inversion model validation of ground emissivity. Contribution to the development of SMOS algorithm

F. Demontoux[1], B. Le Crom[1,2], G. Ruffié[1], JP. Wigneron[2], J. Grant[2,3] D. Medina Hernandez[1]

1 University Bordeaux - Laboratory IMS-UMR 5218 - 16 av.front Pey-Berland 33607 Pessac -France
2 INRA-unit of Bioclimatologie, BP 81, Villenave d' Ornon Cedex 33883 -France
3 Faculty of Earth and Life Sciences, Vrije Universiteit Amsterdam, De Boelelaam 1085- Nederland.
francois.demontoux@ims-bordeaux.fr

*Abstract*— **SMOS (Soil Moisture and Ocean Salinity), is the second mission of "Earth Explorer" to be developed within the program "Living Planet" of the European Space Agency (ESA). This satellite, containing the very first 1.4GHz interferometric radiometer 2D, will carry out the first cartography on a planetary scale of the moisture of the grounds and the salinity of the oceans. The forests are relatively opaque, and the knowledge of moisture remains problematic. The effect of the vegetation can be corrected thanks a simple radiative model. Nevertheless simulations show that the effect of the litter on the emissivity of a system litter + ground is not negligible. Our objective is to highlight the effects of this layer on the total multi layer system. This will make it possible to lead to a simple analytical formulation of a model of litter which can be integrated into the calculation algorithm of SMOS. Radiometer measurements, coupled to dielectric characterizations of samples in laboratory can enable us to characterize the geological structure. The goal of this article is to present the step which we chose to validate this analytical model.**

Keywords : Permittivity, litter, soil, moisture, emissivity model, L-band

## I. INTRODUCTION

SMOS (Soil Moisture and Ocean Salinity), whose launching is planned for the horizon 2007, is the second mission of "Earth Explorer" to be developed within the program "Living Planet" of the European Space Agency (ESA) [1]. About the problems of the cycle of water, the data obtained by SMOS will make it possible to establish a chart of the moisture of continental surfaces and salinity of the surface layer of the oceans. The applications are multiple. On Earth, the retention of water in the ground takes a significant place in the climatic evolution because the moisture of the grounds represents a variable key controlling the exchange of water and thermal energy between the ground and the atmosphere. For the sea, salinity is a parameter which influences the circulation of the water masses in the oceans and involves the formation of climatic phenomena such as El Niño. This satellite, containing the very first 1.4GHz interferometric radiometer 2D, will thus carry out the first cartography on a planetary scale of the moisture of the grounds and the salinity of the oceans. There is a direct relation between the moisture of the ground for depths from 2 to 5 cm, the salinity of the oceans and the emissions of terrestrial origin at the frequency of 1.4 GHz. The forests are present in a majority of the pixels of SMOS in tropical, boreal and moderate zone. The forests are relatively opaque, and the knowledge of moisture remains problematic. The effect of the vegetation (tree and under wood) is characterized by its optical thickness τ and its simple diffusion albedo. Its effect can be corrected thanks a simple radiative model. Nevertheless simulations show that the effect of the litter on the emissivity of a system litter + ground is not negligible. It is thus necessary to set up an analytical model which would make it possible to correct the effect of this (these) additional layer (s) and to validate it. The objective is to highlight the effects of this layer on the total multi layer system. This will make it possible to lead to a simple analytical formulation of a model of litter which can be integrated into the calculation algorithm of SMOS in order to collect information on the moisture of the ground from emissivity measurements. A radiometer is installed on the site of the INRA institute (Bray-33) which allows us to take radiometric measurements on under wood of this site. The model of inversion will give us information to consider the geological structure. Measurements of the electromagnetic behaviour in situ can be taken simultaneously with preceding measurements on this ground using a horn antenna and a network analyzer. These measurements, coupled to dielectric characterizations of samples in laboratory can also enable us to characterize the geological structure. All the results obtained by these methods will be confronted in order to validate the analytical model. The goal of this article is to present this step which we chose to validate the analytical model.

## II. PRESENTATION OF THE MODELS

The medium that we will study in this article is the forest. This system can be represented by a tri-layer system made up with vegetation (trees, under wood), with litter (vegetable remains) and ground.

## A. Correction of the vegetation effect

The algorithm developed within the framework of mission SMOS makes possible to associate the required geophysical value SM (for "Soil Moisture") to the measured radiometric value BT (for Brightness Temperature) thanks to a simplified radiative transfer model called the τ-ω model. This model considered the medium as a double-layered system made up of the vegetation (tree and under wood) and of the ground (ground + litter). Thanks to the simplified approach of the radiative transfer equation which represents the assessment of radiative energy [2] it is possible to obtain a simple expression of the emission of a cover. The radiometer measures a brightness temperature TB which it is possible to associate to the total cover emissivity thanks to the approximation of Rayleigh-Jeans. TB is the sum of outgoing energies and can then be broken up in order to know the emissivity of the ground (ground + litter).

$$T_B = (1-\omega_V)(1-\gamma_V)(1+\gamma_V\Gamma_S)T_V + \gamma_V e T_S \quad (1)$$

Where:
e is the soil emissivity;
$T_S$ is the soil temperature;
$\Gamma_S$ is the soil reflectivity and $\Gamma_V$ is the vegetation reflectivity that will be neglected;
ω is the vegetation single scattering albedo;
$\gamma = e^{-\tau/\cos(\theta)}$ is the attenuation coefficient which calculation is based on the optical depth of the vegetation layer (τ) and on the observation angle (θ).

## B. Correction of the litter effect with the reflectivity model τ-ω-Rr

The principle of this model is nearly the same as the previous model. In this model, the reflectivity of the interface litter/air is not neglected any more. In this case we obtain a formulation in series for each of the various contributions to the temperature of brightness of the system. The extraction of the temperature of brightness of the ground alone is then possible starting from this formulation. The following expression of the global brightness temperature is obtained:

$$T_B = \frac{1-\Gamma_L}{1-\Gamma_L\Gamma_S\gamma}\left[(1+\gamma\Gamma_S)(1-\gamma)(1-\omega)T_L + (1-\Gamma_S)\gamma T_S\right] \quad (2)$$

Where:
the index L refers to the litter and the index S refers to the ground
$\Gamma_S$ and $\Gamma_L$ are respectively the reflectivity at the interface ground/litter and the interface litter/air;
$T_S$ and $T_L$ are respectively the temperatures of the ground and the litter;
$\gamma = e^{-\tau/\cos(\theta)}$ is the attenuation coefficient.

## C. Dielectric measurement of soil and litter.

Changes in the dielectric constant of the soil or the litter, which are mainly due to variations in water content, produce a variation in the soil-litter system emission. The knowledge of the electromagnetic properties of these two materials is thus fundamental when computing the system emissivity.

We have measured the permittivity of soil and litter with a rectangular waveguide by taking into account the heterogeneity of the sample. The study was based on soil and litter samples collected in the coniferous forest of Les Landes, near Bordeaux, France. Particular attention was given to measurement errors and the calculation of permittivity [5].

All dielectric measurements presented in this study were done using a wave-guide technique in an air-conditioned room. This method enabled us to work on samples wide enough to account for the layer heterogeneity. The samples were held inside the guide using a support (100 μm thick Mylar sheet), considered to be quasi-transparent for the electromagnetic waves.

The electromagnetic parameters of the samples were determined using the Nicolson, Ross and Weir method (NRW) [6] for rectangular waveguides. This calculation process is based on reflection and transmission measurements which are well adapted to the network analyzers. The principle of the calculation is based on the fact that introduction of the sample into the guide produces a change of the impedance.

The figures 2 and 3 present results of the measurement for the soil. Taking into account the various errors of measurement and the heterogeneity of the sample we obtain ranges of permittivities.

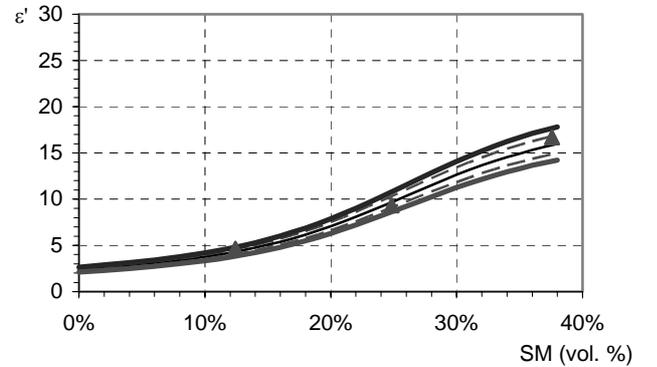

Figure 1. Range of the real part of the soil permittivity.

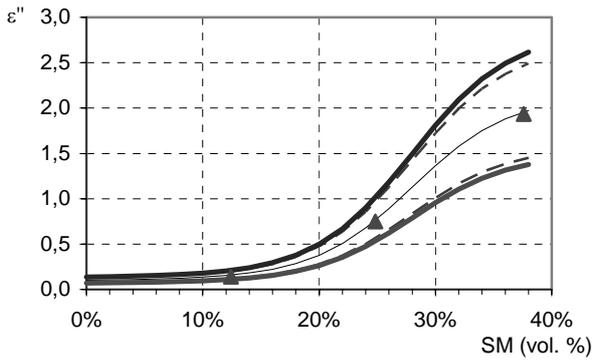

Figure 2. Range of the imaginary part of the soil permittivity.

## III. RESULTS

The first stage was to measure with a horn antenna and a network analyser the in situ electromagnetic response of ground. A model (model HFSS-EM) simulating this experiment was carried out using the software HFSS (High Frequency Structure Simulator) [7].

Comparing the experimental results it was possible for us to improve our knowledge of characteristics of the geological structure profile such as, for example, the thickness of the various layers, their permittivity and thus their moisture. The average thickness of litter was thus evaluated with 4cm and the permittivity of the components of the structure (the day of measurement) refers to a moisture of 25%. The corresponding value of emissivity is 0.895 for the total structure (vegetation+litter+soil).

A radiometric measurement gives the emissivity of the structure. Measurements of emissivity according to moisture were carried out over several months (fig 3). Following figures present curve of emissivity according to the soil moisture. Of course, during the final exploitation of the measurement the soil moisture will be unknown.

To validate the τ−ω−R model, we first applied the τ−ω model on the measurements to correct the effect of the vegetation. Then we obtain the emissivity of the system litter + ground (fig 4).

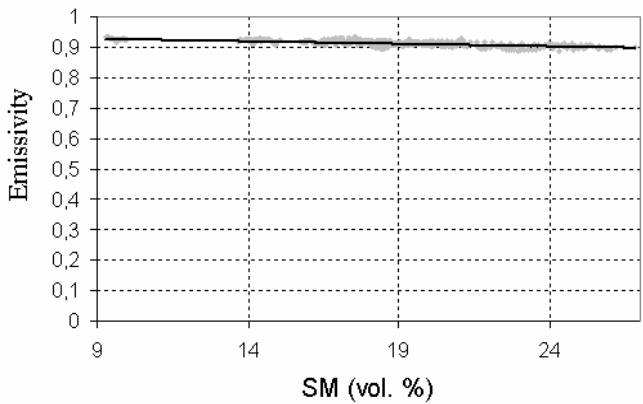

Figure 3. Measurements of emissivity

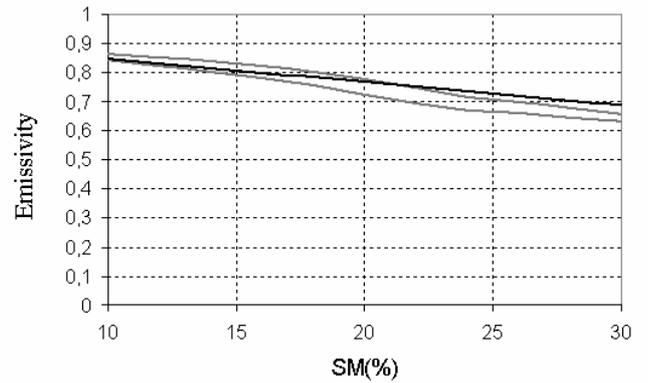

Figure 4. Emissivity of the system ground+litter by the τ-ω method (black) and the HFSS model (grey)

We then determined if the analytical model τ-ω-R allows the computation of the emissivity of the ground when we know the measured emissivity of the complete structure (litter + ground). TS, TL are known while ΓS, ΓL, and, τ and the thickness of the litter are to be determined. For the same input data (measurements of permittivities of the ground and litter according to moisture carried out in laboratory) we compared emissivities of the ground resulting from a HFSS model (HFSS-e model) and the τ-ω-R model. A field of emissivity is defined to take into account the errors to the measures of permittivity introduced into HFSS [4]. The following figure presents the results obtained considering an average thickness of the litter layer of 4 cm with the analytical model as well as the HFSS-e model. On the figure 5, we observe the emissivity of the ground by the τ-ω-R method in black and the emissivity of the ground with the HFSS model in gray according to the moisture of the ground.

The initial value of the emissivity of the complete structure was 0.895. The model τ-ω associates a corrected emissivity of the system ground + litter of 0.72. The exploitation of the model τ-ω-R associates a corrected emissivity of the ground of 0,62 with a moisture of 24% which is a good approximation of the value awaited following measurements in-situ (25%).

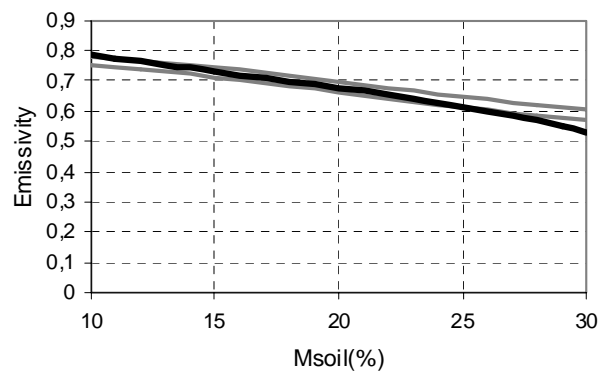

Figure 5. Emissivity of the ground by the τ-ω-R method

## IV. CONCLUSIONS AND PERSPECTIVES

Ours first results confirm the relevance of the τ-ω-R model to correct the effect of the litter. The good agreement of the results resulting from various models consolidated our choice. The next stage of our work which we wish to present will consist in confirming these results by carrying out new series of in situ measurements (radiometric and electromagnetic behaviour). Thereafter we will study the influence of the surface roughness of the ground on emissivity. Accordingly, the influence of the litter coupled with the roughness of the ground will be approached. Indeed, studies showed that terms can be introduced to compute reflectivity of a surface to take into account its roughness. Corrective terms could thus be integrated into the τ-ω-R model to obtain the emissivity of the structure versus the roughness of the soil layer. Those terms will be determinate and validate thanks series of computation using the τ-ω-R model and the HFSS model. In addition to the roughness or the thickness of litter, we will introduce gradients of moisture in order to study equivalent emissivity.